# Depth-dependent interplay of dynamical heterogeneity and chain dynamics at the surface of glass-forming polymers

Bao T. Ma[a], David S. Simmons[a*]

[a]Department of Chemical, Biological, and Materials Engineering, The University of South Florida, Tampa, Florida

[*]dssimmons@usf.edu

**Abstract**: Polymer thin films exhibit pronounced interfacial mobility gradients that modify chain relaxation, yet how these gradients govern chain-scale dynamics across depth remains incompletely understood. Using molecular dynamics simulations of freestanding glass-forming polymer films, we resolve how depth-dependent variations in segmental relaxation shape chain dynamics across a wide range of displacement scales. Near the free surface, accelerated segmental mobility suppresses Rouse-regime scaling exponents to values as low as $\gamma \cong 0.4$, reflecting transient localization induced by interfacial mobility gradients rather than topological entanglement. In contrast, the film interior exhibits enhanced Rouse scaling exponents consistent with predictions of the Heterogeneous Rouse Model (HRM), indicating that bulk dynamic heterogeneity compresses the Rouse regime. Mapping the minimum scaling exponent $\gamma_{min}$ across depth reveals a linear gradient that separates the bulk-like enhancement regime from the surface-induced suppression regime of chain dynamical scaling. Together, these results demonstrate that bulk and interfacial dynamic heterogeneity modify chain relaxation in opposite ways and establish Rouse scaling as a sensitive, spatially resolved probe of glassy dynamical heterogeneity and interfacial dynamical gradients in polymers.

## Introduction

When glass-forming liquids are cooled towards their glass transition temperature $T_g$, they exhibit an onset of dynamic heterogeneity – spatial variations in the local rate at which molecules move and relax.[1–13] In the bulk, these heterogeneities are spatially correlated but not spatially organized; they take the form of interspersed slow and fast domains. In the nanoscale vicinity of interfaces, a new and even more dramatic form of dynamic heterogeneity emerges – enormous, spatially organized gradients in relaxation time in the direction normal to the surface.[14–23] In polymers, the presence of these two types of dynamic heterogeneity raises both fundamental and practical questions regarding how polymers move, relax, and deform in their glass formation range – both in the bulk and near interfaces. Foundational models treat polymer chain motion and viscosity as emerging from a hierarchical process of increasingly large sections of chains relaxing within a uniform environment of segmental dynamics.[24,25] How does this picture change in the presence of an enormous distribution of local segmental dynamics? How does overall chain motion occur when a polymer chain spans domains with highly differing local segmental relaxation times (or, equivalently, friction coefficients)? How does the interplay of bulk and surface-like dynamic heterogeneities impact polymer chain motion near interfaces?

In the bulk, recent work by our group[26,27] has indicated that dynamic heterogeneity drives a long-observed[28–45] breakdown in Time-Temperature-Superposition (TTS) near $T_g$. This TTS breakdown takes the form of altered, temperature dependent Rouse scaling exponents in the glass formation range.[26,27] The Heterogeneous Rouse Model (HRM) predicts this behavior by generalizing the Rouse theory to account for bulk dynamic heterogeneity.[27] Notably, this model is successful in tests against simulations and experiments[26,27] even though it does not account for spatial correlations within dynamic heterogeneity; i.e. the HRM incorporates a distribution of segmental mobilities but treats this distribution to be spatially random. This success is likely due to the relatively short range and relatively disordered nature of spatial correlations in bulk dynamic heterogeneity. Altered Rouse scaling, in the form of enhanced scaling exponents for the dependence of relaxation modulus or complex modulus on time or frequency, then emerges within this theory, not from spatial correlations but from simple averaging over the distribution of segmental mobilities. With increasing strand length, chain dynamics interpolate from sampling the forward moment to one over the inverse moment of the distribution of segmental friction coefficients. This leads to a prediction that bulk dynamic heterogeneity increases effective Rouse scaling exponents relative to ideal Rouse expectations[26,27]. In particular, the Rouse theory predicts that for times (frequencies) between the segmental relaxation and terminal relaxation (in unentangled chains) or onset of entanglement (in entangled chains), polymers should obey the following scaling relations with scaling exponent $\gamma = ½$.

$$\langle r^2 \rangle \propto t^{\gamma} \qquad (1)$$

$$G(t) \propto t^{-\gamma} \qquad (2)$$

$$G(\omega) \propto \omega^{\gamma} \qquad (3)$$

The HRM, in contrast, predicts and explains the empirical observation[26,28,45] that $\gamma > ½$ and grows on cooling, as the breadth of the distribution of local mobilities (magnitude of dynamic heterogeneity) grows.

This scenario of simple, spatially uncorrelated averaging effects becomes conceptually implausible near interfaces. Near free surfaces, for example, the segmental mobility of polymers and other glass-forming liquids can be enhanced by many orders of magnitude. The basic form and likely underlying physical origin of these gradients, at least in equilibrium at timescales accessible to simulation, have become reasonably resolved over the last decade.[14,23,46–49] The current theory that provides the best empirical description of these results predicts that these gradients emerge because a free surface weakens local caging constraints on particles near the surface, while also truncating longer-ranged elastic barriers to relaxation that emerge upon approach to $T_g$.[49] These effects lead to a gradient in $\log(\tau_\alpha)$ (where $\tau_\alpha$ is the segmental relaxation time) that initially decays into the film in an exponential manner (decay lengthscale of several nm or segmental diameters), with a weaker power-law tail extending much deeper into the material.[50]

For the purposes of this investigation, the initially sharp exponential gradient raises serious questions regarding the nature of polymer chain motion and relaxation near surfaces. When a polymer chain spans this high magnitude, medium range gradient, how does the overall chain move over time? The answer to this question is of central significance to the understanding of rheology, flow, and mechanical response in interfacially-rich polymers.

Considerable experimental effort has been brought to bear on developing methods capable of probing nanoscale polymer surface dynamics or mechanics with sufficient resolution to understand polymer deformation in this region. Approaches have included measurement of recovery of step changes in the height of a film,[22] measurement of surface-stress induced deformation of a film's leading edge,[21] and XPCS measurements of film surface fluctuations,[51] among others. These studies have clearly established that surface gradients in $T_g$ and segmental relaxation time indeed lead to alterations in polymer chain motion and viscosity near film surfaces. A full resolution of the question of the precise nature and mechanism of these changes, however, demands time- or frequency-resolved response data capable of probing changes in the *spectrum* of dynamic and rheological responses at the film surface.

Recently, new experiments and simulations have probed time-resolved creep and molecular displacement data at film surfaces to provide an initial glimpse at this modified surface response spectrum. These findings revealed that the surface of polymer glasses behave as transiently crosslinked polymers, even at subentangled molecular weights.[52] Strikingly, while surface gradients accelerate *segmental* relaxation rates, they simultaneously lead to emergence of entanglement-like plateaus in *whole-chain* dynamics and rheological response. This indicates that interfacial effects qualitatively alter the physical mechanisms governing polymer relaxation, reshaping how heterogeneity couples to chain motion. In our prior work, we developed a theory for this effect, wherein this plateau behavior is predicted to arise from covalent tethering of surface segments to more deeply-placed and slower-moving interior segments.[52]

From a formal mechanistic standpoint, that theory (consistent with experimental findings) suggests that that the spatially-organized dynamic heterogeneity (mobility gradient) present at a film surfaces *suppresses* the scaling exponents for chain displacement vs time in the Rouse regime of polymer intermolecular dynamics.[52] Notably, this effect is precisely opposite to that observed and predicted in the bulk, where dynamic heterogeneity enhances scaling exponents in the Rouse regime.[26,27] As noted above, in the bulk, dynamic heterogeneity alters scaling in the Rouse regime such that $1/2 < \gamma < 1$, with $\gamma$ increasing on cooling as dynamic heterogeneity grows in magnitude.[26,27] In contrast, our recent work indicates that at the film surface, $\gamma < 1/2$, drops on cooling, and can drop to as low as ¼, comparable to chain displacement scaling in the reptation regime for entangled systems.[52]

However, the theoretical understanding developed in our prior work for these effects was limited to the segmental monolayer at the surface of the film. This leaves a core set of open questions. How do heterogeneity effects on chain dynamics evolve with depth in a polymer film? In particular, what is the spatial extent of the crossover from surface-like suppression of Rouse scaling to bulk-like enhancement? *The answer to this question will determine the thickness of the rheological interphase at the surface of a film – a property central to practical applications such as film processing and polymer sintering.* Moreover, how does the interplay between surface mobility gradients and bulk dynamic heterogeneity shape polymer motion near interfaces? Given that bulk and dynamic heterogeneity evidently modify Rouse scaling in opposing manners, how can we understand the crossover between these two qualitatively distinct regimes of heterogeneity effects on polymer chain motion? Understanding this crossover is essential for developing a unified picture of relaxation in confined polymer systems such as polymer films and nanocomposites, and for designing polymeric materials with tailored interfacial properties and rheological response.

Here, we investigate the depth-resolved dynamics of polymer chains in freestanding films using molecular dynamics simulations. By analyzing chains' mean square displacement and associated scaling exponent $\gamma$ as a function of depth in the film, we uncover a rich gradient in dynamical behavior that evolves with temperature and molecular weight. Our results show that while chain segments near the surface undergo larger displacements, their scaling exponents in the Rouse regime are suppressed, indicating a transient tethering effect. In contrast, interior chains exhibit elevated scaling exponents consistent with HRM predictions, reflecting the influence of bulk dynamic heterogeneity. Moreover, we find that the crossover between these regimes occurs over a domain of order nanometers in thickness, with the range of this gradient varying weakly with molecular weight. Our findings further indicate that surface gradient tethering effects are distinct from topological entanglement effects, occurring at smaller displacements and shorter timescales. These findings reveal a rich dynamical and mechanical environment near the surface of glassy polymer films, emerging from an interplay of glassy physics and chain connectivity. Moreover, they suggest that polymer chain motion may provide a novel and uniquely sensitive way of probing dynamic heterogeneity near surfaces and in the bulk.

# Methods

## *Simulation Methodology*

This work presents additional analysis of simulations previously reported in *Nature*[52]; full simulation methodology details are available in the original publication. A summary is provided here for convenience. Simulations are conducted using the Large-scale Atomic/Molecular Massively Parallel Simulator (LAMMPS),[53]. Our simulations employ a bead-spring polymer model extended from the foundational work of Kremer and Grest.[54] This approach incorporates attractive forces among non-bonded beads, which interact using a 12-6 Lennard-Jones (LJ) potential,

$$E_{ij} = 4\varepsilon\left[\left(\frac{\sigma}{r}\right)^{12} - \left(\frac{\sigma}{r}\right)^{6}\right], \qquad r < r_{cut} \qquad (4)$$

with energy parameter $\varepsilon$ equal to 1 and range parameter $\sigma$ equal to 1 $\sigma_{LJ}$ (where $\sigma_{LJ}$ is the Lennard Jones distance unit and corresponds roughly 1 nm[55,56]), and $r_{cut}$ = 2.5 $\sigma_{LJ}$. Bonded interactions are modeled via a combination of the

Finitely Extensible Nonlinear Elastic (FENE) potential with the 12-6 LJ potential,

$$E_{FENE} = -0.5 R_0^2 \ln\left[1 - \left(\frac{r}{R_0}\right)^2\right] \\ + 4\varepsilon_{FENE}\left[\left(\frac{\sigma_{FENE}}{r}\right)^{12} - \left(\frac{\sigma_{FENE}}{r}\right)^6\right] \quad (5)$$

with $K = 30$, $R_0 = 1.3$, $\varepsilon_{FENE} = 1$, and $\sigma_{FENE} = 0.8$. In equation (5), the 12-6 LJ term is cut off at $r_{cut} = 2^{1/6}\sigma_{LJ}$. These parameters are modified from the most commonly employed FENE parameters in a manner that enhances resistance to crystallization near interfaces.[57] This modeling approach has been validated in earlier work on confined polymer systems,[58,59] and comparable potentials serve as the basis for numerous studies of polymer glass transition dynamics near interfaces and under confinement.[55,60–69]

After generating random initial configurations, our simulation protocol initiates by isothermally annealing the system at a reduced LJ temperature of 1.5 for $5 \times 10^5 \tau_{LJ}$, where $\tau_{LJ}$ denotes the LJ unit of time. During this phase, we maintain a constant box size that is considerably larger than what the equilibrium density of the polymer would ordinarily require. This setup facilitates the rapid consolidation of the polymer into a film, which then undergoes an efficient thermal annealing process. We employ a Nose–Hoover thermostat to control temperature, with a damping parameter of $2\ \tau_{LJ}$ for temperature control.

Our simulations' thermal protocol then follows the Predictive Stepwise Quench (PreSQ) algorithm described in prior studies[60] to achieve equilibrium over a range of temperatures approaching $T_g$. This approach begins by annealing the system at temperatures well above the typical glass transition. In subsequent quenching phases, temperatures are lowered in stages, followed by isothermally annealing to equilibrium. During each stage, we analyze the segmental relaxation dynamics based on the self-part of the intermediate scattering function computed at a wavenumber comparable to the first peak in the structure factor. The system is considered to be in equilibrium at a given temperature if its segmental relaxation time is no more than one tenth the isothermal annealing period performed at that temperature.

## Simulation Analysis

Simulation analysis employs the Amorphous Molecular Dynamics Analysis Toolkit (AMDAT).[70,71] We analyze segmental and chain dynamics based on the segmental mean square displacement $<r^2(t)>$, defined as

$$\langle r^2(t) \rangle = \frac{1}{S}\sum_j^S \frac{1}{N}\sum_i^N \left[r_i(s_j + t) - r_i(s_j)\right]^2, \quad (6)$$

where $s_j$ is the initial time of observation, $t$ is the duration of displacement, $S$ is the count of initial times $s_j$ over which the mean square displacement $<r^2(t)>$ is averaged, $r_i(s)$ is the position of particle $i$ at time $s$, and $N$ is the number of particles.

We quantify segmental dynamics by partitioning each film into equally thick surface bins based on distance from the free surface ($i = 0$) toward the interior (larger $i$). Each bin is of thickness 1 $\sigma$ (1 particle diameter). Each bin is therefore centered at a position $z = 0.5 + i$ segmental diameters.

For each depth bin $i$, we compute the mean-squared displacement (MSD) $<r^2(t)>_i$. To exploit the mirror symmetry of the two free interfaces, we average partner bins located at equal distances from the opposite surfaces ("top" and "bottom") to obtain a single MSD profile per depth. As described in Results, the apparent exponent $\gamma(t)$ within any bin and any time is the logarithmic slope of $<r^2(t)>$ and is obtained by numerical differentiation on the logarithmic time axis using centered finite differences. We examine $\gamma$ both as a function of time and, more usefully for cross-state comparisons, as a function of $<r^2(t)>$, which fixes the spatial scale.

## Results

### Time-resolved scaling exponents

To examine how free surfaces influence polymer chain dynamics, we analyze the mean square displacement (MSD) $<r^2(t)>$ and its scaling behavior across different depths and temperatures. We begin by computing the time-dependent MSD of chain segments within spatial bins ranging from the surface to the interior of the film. Figure 1 presents log-log plots of $<r^2(t)>$ versus time at both high and low temperatures. At high temperature (T = 1.5), the MSD curves from all depths closely overlap, indicating that dynamics do not appreciably differ as a function of depth in the film. Upon cooling, however, a pronounced enhancement in mobility emerges near the surface. Segments near the free surface exhibit significantly larger $<r^2(t)>$ at any given time than do segments in the interior, reflecting enhanced surface segmental mobility. The emergence of this gradient as temperature decreases is expected from the literature of interfacial alterations in dynamics, where these alterations are found to emerge and grow in strength upon cooling towards $T_g$.

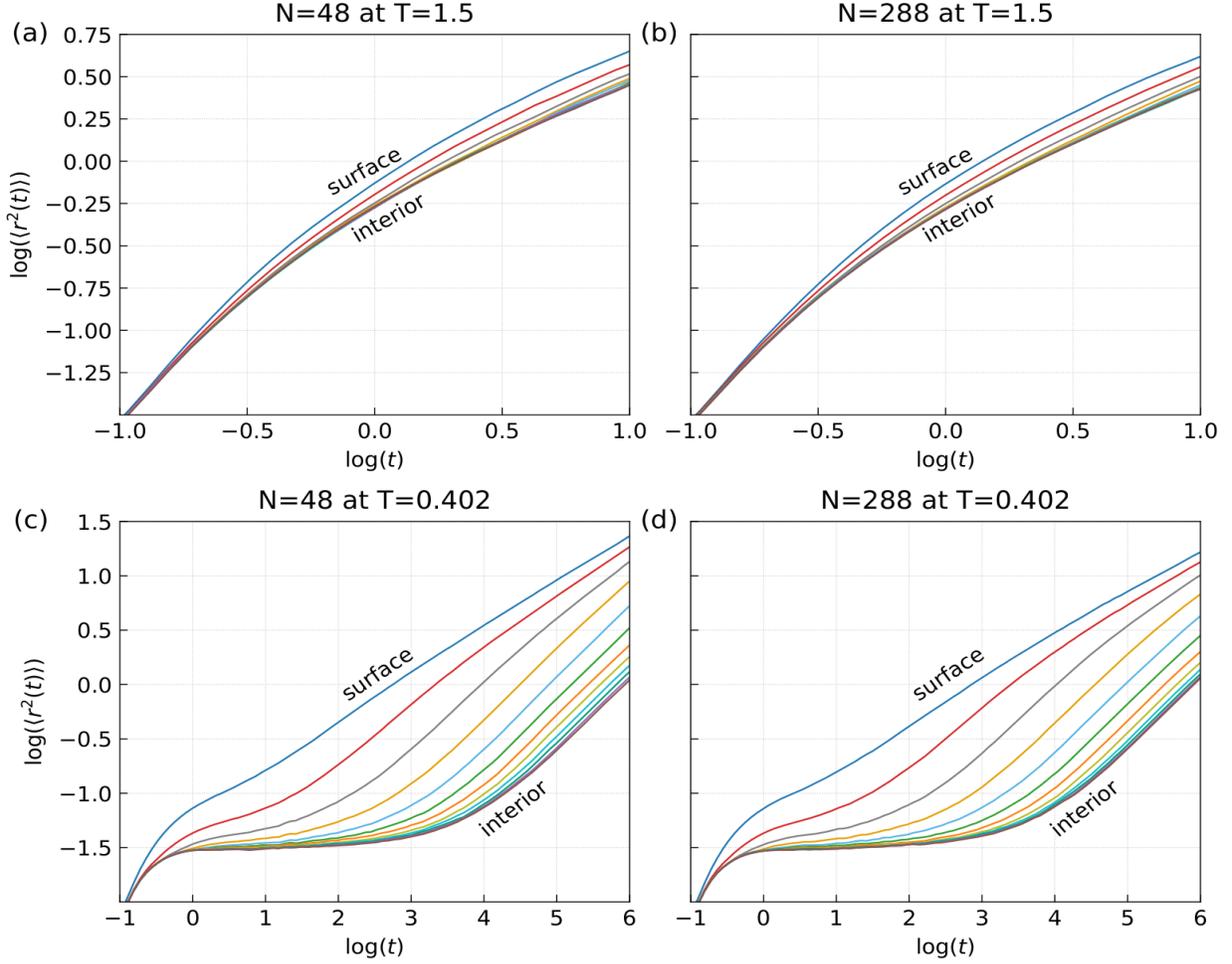

*Figure 1. Depth-resolved mean square displacement of polymer segments at high temperature [(a) and (b)] and low temperature [(c) and (d)]. Each curve, from top to bottom of each panel, represents a sequentially deeper bin from the surface, beginning with bin index i = 0 at the top of each panel. Each bin is centered at a position z = 0.5 + i segmental diameters from the free surface.*

To gain further insight into the mechanisms governing displacement as a function of time and position in the film we obtain an effective value of $\gamma$ as the logarithmic slope of $<r^2(t)>$ via numerical differentiation,

$$\gamma_{\mathit{eff}} = \frac{d\log\langle r^2(t)\rangle}{d\log t} \qquad (7)$$

which provides the instantaneous apparent scaling exponent of displacement with time (Figure 2). We emphasize that $\gamma_{\mathit{eff}}$ obtained in this manner is an *effective* scaling exponent, since equation (7) neglects a term proportional to $d\gamma/d\log t$. This is appropriate in regimes in which $\gamma$ does not evolve rapidly with logarithmic time. Since our focus is on regimes in which $\gamma$ exhibits a plateau or soft minimum, this is appropriate to the purposes of this analysis, and we henceforth refer to $\gamma_{\mathit{eff}}$ obtained in this manner simply as $\gamma$ for simplicity.

Under these conditions, $\gamma$ then provides insight into the physical regime of motion: ballistic-like ($\gamma = 2$), diffusive-like ($\gamma = 1$), Rouse-like ($\gamma \sim 0.5$), and reptation-like ($\gamma \sim 0.25$ for ideal reptation, or $\gamma \sim 0.4$ with chain end effects). When $\gamma$ falls below 0.5, it typically points to some mechanism of chain localization, either from segmental caging in glassy polymers, chain entanglement in entangled bulk polymers, or transient crosslinks in physical networks.[24,55,72–74] We have also shown in prior work that sub-0.5 scaling can emerge from localization by surface dynamical gradient effects.[52] However, this observation raises a key question: how can we distinguish among these mechanisms when $\gamma$ deviates from classical Rouse-like behavior? For instance, when $\gamma < 0.5$, is the suppression attributable to caging, entanglement, or surface gradient effects? Conversely, when $\gamma > 0.5$, what physical process drives this enhancement? These questions motivate a more detailed analysis of $\gamma$ as a function of displacement rather than time, since distinct localization mechanisms tend to manifest at characteristic displacement scales. This approach enables us to disentangle the physical origins of suppressed or enhanced scaling behavior by linking each $\gamma$ regime to its characteristic displacement scale across spatial depths and thermal conditions.

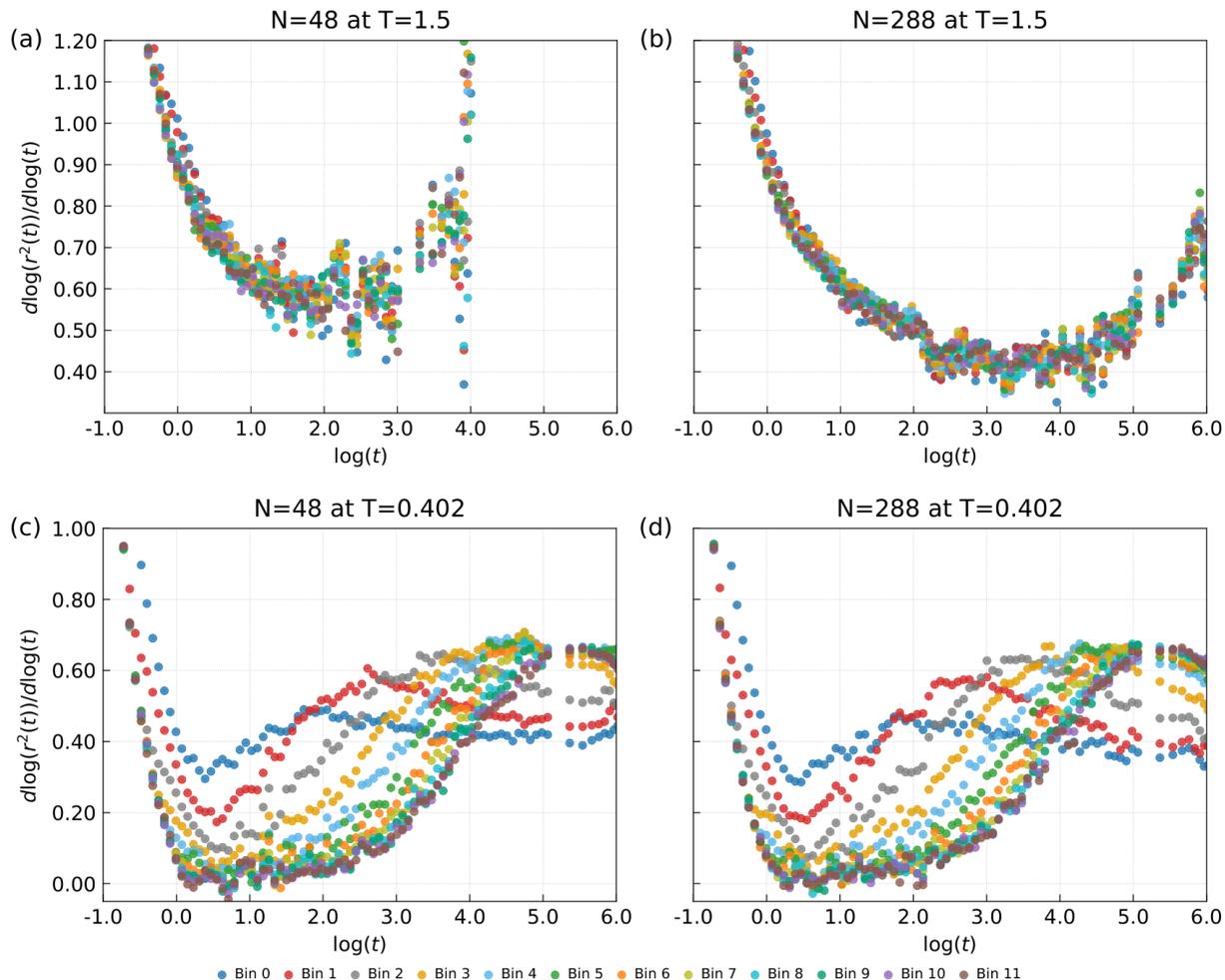

*Figure 2. Instantaneous scaling exponent γ, derived from MSD as per eq. (7), as a function of log(time) for polymer beads located in bins of index i shown in the legend. Each bin is centered at a position z = 0.5 + i segmental diameters from the free surface. The chain length N and temperature T corresponding to the data in each panel are noted at the top of each panel. Unentangled polymers are on the left and entangled chains are on the right; high temperature data are shown in the upper panels and low temperature data in the lower panels.*

As shown in Figure 2(a) and (b), at high temperature (T = 1.5), the scaling exponent $\gamma$ remains consistent across all depths, confirming uniform dynamics across the film. At low temperature (Figure 2 2(c) and 2(d)), however, $\gamma(t)$ differs significantly between surface and interior bins. Moreover, the low temperature systems exhibit a curious crossover behavior. At short times, interior bins exhibit $\gamma$ approaching zero, corresponding to strong glassy particle segmental caging, while surface particles exhibit much weaker caging behavior (higher $\gamma$). However, at longer times the scaling exponents for surface and interior particles cross over, such that at long times surface particles exhibit lower $\gamma$ than do interior particles.

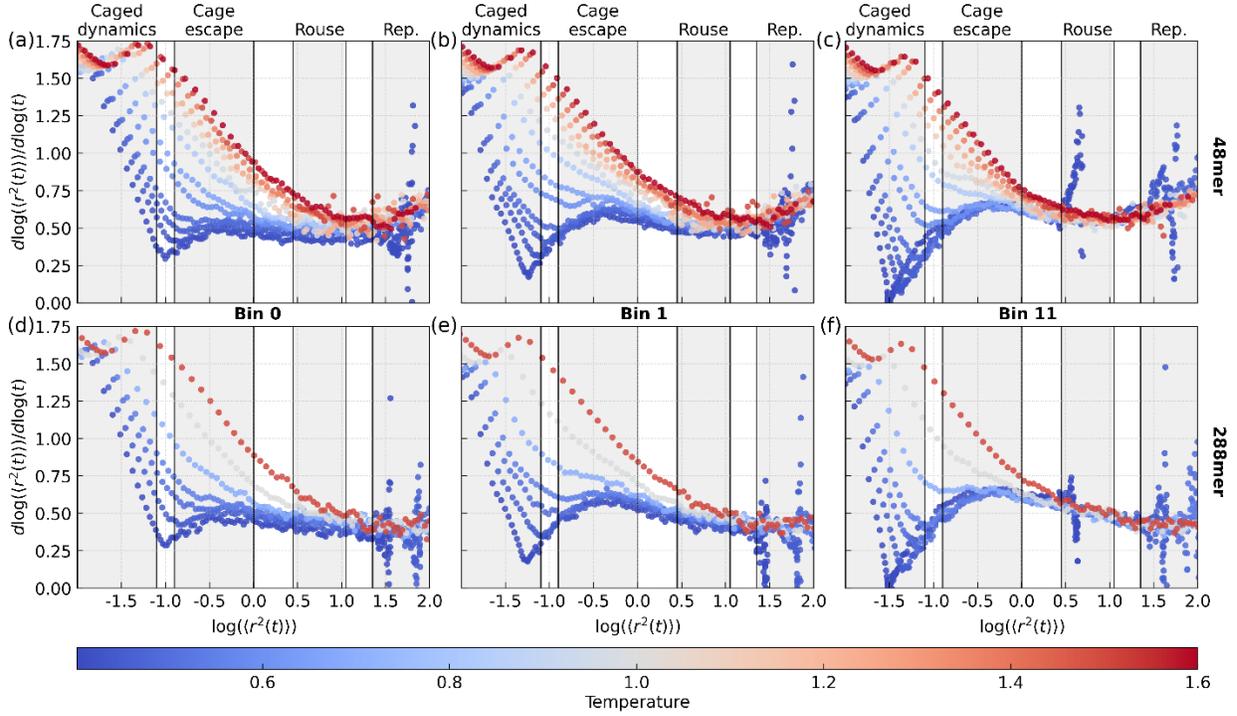

*Figure 3 Scaling exponent γ of segments of polymers with unentangled (N=48) [(a), (b) and (c)] and entangled (N=288) [(d), (e) and (f)] chain lengths, plotted vs logarithmic mean square displacement, across a range of temperatures indicated by the color bar. Each panel corresponds to particles in a distinct spatial bin i, noted in the middle of each column. Each bin is centered at a position z = 0.5 + i segmental diameters from the free surface.*

## *Relaxation mechanisms in the bulk-like midfilm*

To better understand this behavior, we replot the scaling exponent $\gamma$ as a function of mean squared displacement $<r^2>$ in Figure 3. We begin by considering the temperature-dependent variation of $\gamma(<r^2>)$ in the bulk-like film interior, seen in Figure 3c and f. As can be seen here, upon cooling towards $T_g$ the system exhibits an onset of caged rattling behavior, with $\gamma$ dropping towards 0, in the displacement range of $<r^2> \cong 10^{-2}\ \sigma_{LJ}^2$ to $10^{-1}\ \sigma_{LJ}^2$. This is a core signature of the glass transition – arrested cage dynamics on displacement length scales appreciably less than the particle or segment diameter.

As this caged behavior emerges, a cage escape regime likewise emerges in the displacement range of $<r^2> \cong 10^{-1}\ \sigma^2$ to $10^{-0}\ \sigma^2$. This takes the form of a transient peak in $\gamma$ on length scales modestly less than the bead size. We assign this to particle hopping, given that it corresponds roughly with the known (approximately bead-scale) scale of particle hops in deeply supercooled liquids, and given that it evidently represents a transient regime of higher mobility. Signatures of hopping behavior have previously been reported in other dynamic correlation functions such as the self-part of the van Hove function,[75] where it sometimes appears as an emergent bimodal distribution (or at least excess tail) in the distribution of particle displacements at times comparable to the segmental relaxation time. We are unaware of any report of this length scale being extracted directly from the mean-square displacement. This may provide a complementary manner of identifying the onset temperature of pronounced hopping along with the length scales and time scales of hop events (based on the emergence and location of a peak in $\gamma$ vs $<r^2>$).

At larger displacement scales, these bulklike $\gamma$ vs $<r^2>$ data exhibit a transition from glass-physics-dominated dynamics to chain-connectivity-dominated dynamics. In the displacement range of $<r^2> \cong 10^{0.45}\,\sigma^2$ to $10^{1.05}\,\sigma^2$, the subentangled system exhibits a reduction of $\gamma$ from its peak value of ~ 0.7 in the hopping peak to a value in the range of 0.5 to 0.65 in the range of $\log(<r^2>)$ from 0.5 to 1.0 This range of $<r^2>$ is consistent with the displacement scale expected for Rouse motion: both the square gyration radius of a 48-mer and the square gyration radius of the entanglement strand for this system are of order $10^1$; the Rouse regime is expected to initiate after the segmental relaxation event and then persist up to the approximate scale of $<r^2> \sim <R_g^2> \sim 10^1$. Moreover, the $\gamma$ values of 0.5 to 0.65 are consistent with the expectations of the Heterogeneous Rouse Model for Rouse-like chain motion in a dynamically heterogeneous bulk environment. Predictions of the HRM are also consistent with the relatively weak temperature dependence of $\gamma$ in this displacement regime.[27] Here, the temperature dependence of $\gamma$ is controlled by the temperature dependence of dynamic heterogeneity rather than by the temperature dependence of the caging strength (as in the segmental caging regime) – the former is much weaker than the latter. In essence, highly temperature dependent glassy dynamics govern relaxation on shorter lengths-scales; on chain scales, they modulate relaxation but do not alter its basic character.

Finally, for the entangled ($N$ = 288) system only, dynamics cross over into a regime with $\gamma \cong 0.4$ for the longest displacement scales probed ($<r^2>$ > $10^{1.5}$). This reflects a crossover to entangled chain reptation behavior. The location of this crossover is reasonable given the reported entanglement molecular weight of this model, approximately 84 repeat units.[76] This corresponds to a square gyration radius of the entanglement strand in the range of $10^1 – 10^{1.5}$, anticipating that chain dynamics must be constrained by entanglements on displacement scales larger than this.

This representation of the data thus allows us to associate value of $\gamma$ with particular length scales of displacement and thus to more clearly assign a physical mechanism to each dynamical regime. Displacements on length scales much less than the segment size (say in the range of 0.1 $\sigma^2$ ≪ 1 $\sigma^2$) typically correspond to cage rattling, where segments are temporarily trapped by their neighbors. As $<r^2>$ increases to approximately 0.3 to 1.0, we enter the regime of segmental relaxation, where segments begin to escape local cages and undergo cooperative motion. At larger displacements in the range of $10^0$ to $10^1$, Rouse-like motion of chain modes emerges, characterized by subdiffusive scaling ($\gamma \cong 0.5$) in unentangled chains. In long, entangled chains, a plateau in $\gamma$ (value near 0.4) emerges on larger length scales beyond the scale of entanglement, indicating the onset of reptation or entanglement-dominated dynamics.

## *Interfacial alterations in chain dynamics*

With this bulk-like midfilm baseline in hand, we now turn to the question of how the free surface alters chain dynamics. As in the bulk, segmental caging and hopping behaviors at low particle displacement emerge on cooling at the surface (Figure 3 (a), (b), (d), (e)). However, these phenomena are weakened as compared to the more bulk-like midfilm. This is a direct reflection of the accelerated segmental dynamics and weaker segmental caging that emerges at the surfaces of glass-forming liquids.[14,18,62,77–79] At larger displacements, however, proximity to the surface has a qualitative effect on chain scale dynamics, leading to emergence of a more pronounced temperature dependence to $\gamma$ in the Rouse regime. This effect can be seen in a complementary manner in Figure 4, where we plot $\gamma(<r^2>)$ at fixed temperature over a range of depths in the film in a single panel. At high temperatures, chain dynamics are essentially invariant through the film. At low temperature, the scaling exponent $\gamma$ in the Rouse regime dramatically drops, from 0.6 or greater in the bulk to ~ 0.4 near the film surface.

This behavior marks a crossover from the behavior characteristic of bulk dynamic heterogeneity. The sub-Rouse scaling observed near the surface is not due to entanglement, as the associated displacements are far smaller than those required for topological entanglement constraints. In the N = 288 system, genuine reptation behavior only appears at much larger displacements (Figure 3(f), Figure 4(d)). Moreover, entanglement density tends to decrease, not increase, near film surfaces and under confinement,[80] and this was confirmed for this system in our prior work.[81] This further confirms that the surface suppression of $\gamma$ arises from dynamical gradients rather than entanglement. This surface suppression in $\gamma$ is a direct consequence of the surface gradient tethering effect reported in our prior publication.[81] Our new results make clear that this effect is entirely distinct from topological entanglement, as it operates on chain displacements far lower than those associated with topological entanglement in this model.

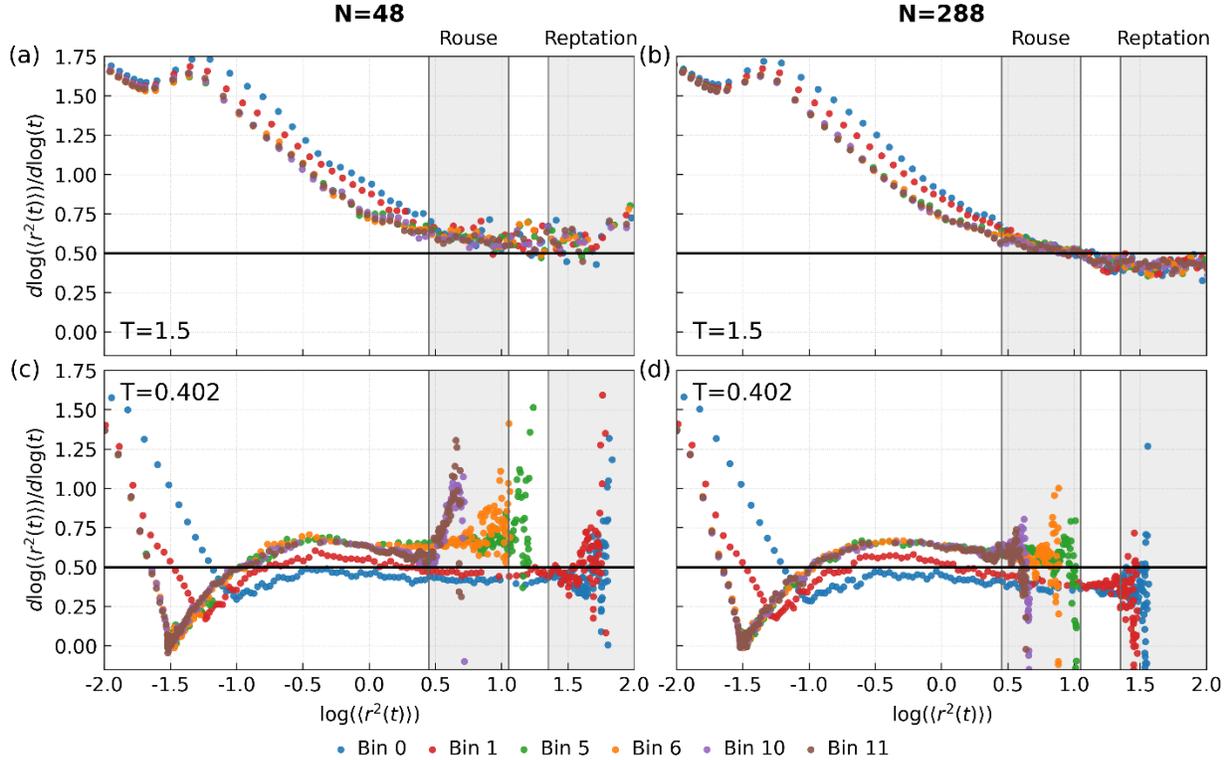

*Figure 4. Displacement-dependent scaling exponent γ of polymer chains plotted vs. logarithmic mean square displacement for chains in spatial bins noted in the legend. Each bin is centered at a position z = 0.5 + i segmental diameters from the free surface. The chain length N and temperature T corresponding to the data in each panel is noted at the top of the panel. Unentangled polymers are on the left and entangled chains are on the right; high temperature data are shown in the upper panels and low temperature data in the lower panels.*

We are unable to simulate long enough times (and thus large enough displacements) at low temperatures to determine whether this entanglement-like surface-tethering region transitions smoothly into the topological entanglement regime at larger displacements, or whether there is a transitional regime of higher mobility separating these two highly constrained regimes. The answer to this question may be nonuniversal.

We next extract a minimum value of $\gamma$ that is characteristic of chain (rather than segmental) motion for each bin and temperature (Figure 5). This enables us to quantify the gradient in chain displacement physics as a function of distance from the film surface. To do so, for each temperature and bin we identify the minimum value of $\gamma$ within the chain dynamics regime of displacements. We first fit $\gamma(\langle r^2 \rangle)$ using a cubic smoothing spline[82–86] with a spline smoothing parameter s $\cong$ 0.08, which we find preserves broad features while suppressing point-to-point noise. We then truncate this spline to search only the region above $\langle r^2 \rangle = 10^{-0.5}$ (thus omitting the segmental-scale caging regime). We further truncate the last 0.45 decades of data on the $\langle r^2 \rangle$ axis, as this regime is generally excessively noisy due to weaker sampling. Within the remaining target region, we find the minimum value of $\gamma$, representing the condition of strongest chain localization (sub-diffusion). We then examine, in Figure 5, the spatial variation of the scaling exponent for most localized chain dynamics $\gamma_{min}$ as a function of distance z from the free surface.

As shown in Figure 5, for both simulated unentangled and entangled systems, the $\gamma_{min}(z)$ gradient grows in magnitude on cooling and has an approximately linear form. This is qualitatively distinct from the form of the underlying relaxation time gradient, which obeys an exponential spatial variation of $\log(\tau)$. This indicates that the interplay of the $\tau$ gradient with chain connectivity effects qualitatively alters the gradient at the level of whole-chain dynamics.

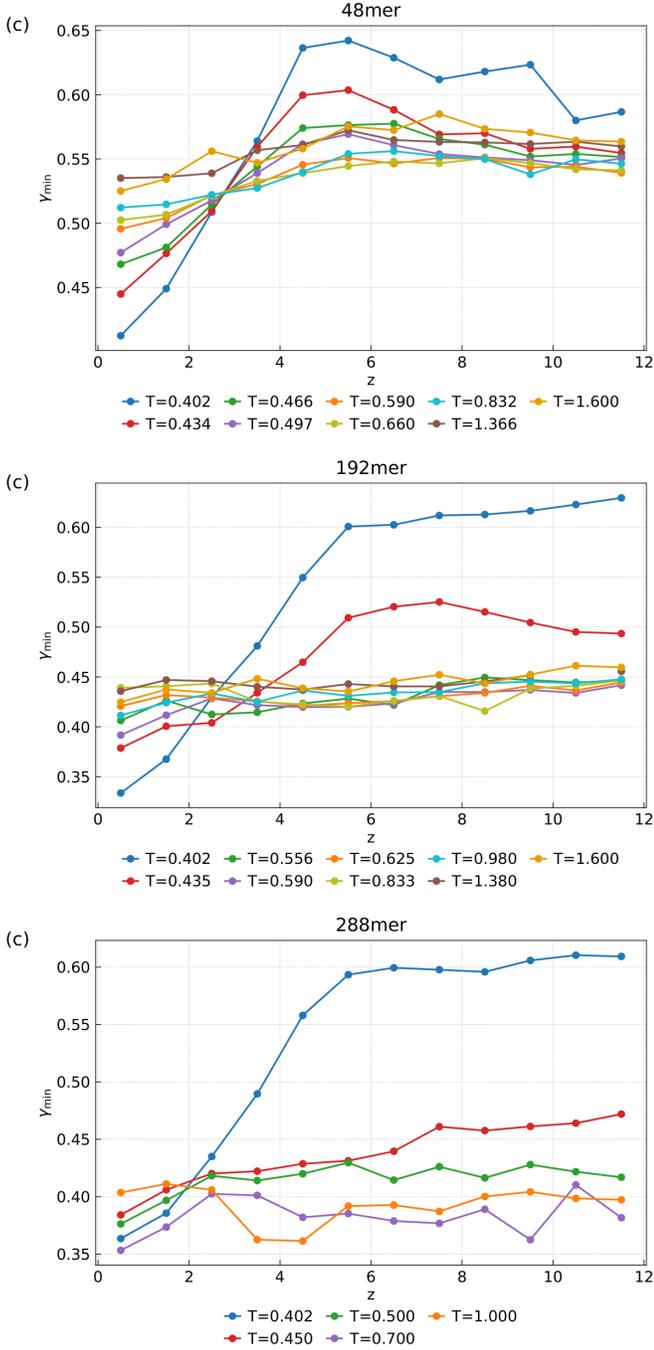

*Figure 5. Spatial gradient of minimum chain dynamics scaling exponent $\gamma_{min}$ as a function of distance z from the film surface. Distinct datasets in each panel correspond to reduced Lennard-Jones temperatures shown in the legends. The panels correspond to distinct chain lengths of (a) 48 beads, (b) 192 beads, and (c) 288 beads.*

The magnitude of the gradient in chain dynamics grows with increasing chain length, with longer chains exhibiting lower surface values of $\gamma$. This reflects stronger pinning effects near the interface in longer chains, which on average penetrate more deeply into the film and thus pass through slower-relaxing regions. Moreover, in longer chains the surface-induced suppression in $\gamma$ penetrates modestly further into the film. However, we note that for a chain with 288 segments, the mean end-to-end distance is expected to be in the range of 15-20$\sigma$. As shown by Figure 5c, even in this molecular weight range the surface gradient in $\gamma$ does not dramatically exceed in range the shorter-ranged spatial gradient in segmental relaxation time. This suggests that surface effects on chain dynamics likely saturate once chains are long enough to fully span the distance from the accelerated surface to the bulk-like interior segmental dynamics environments.

These gradients in $\gamma_{min}$ reflect how different dynamical modes sample the spatial heterogeneity of the film. Near the surface, high-frequency modes are dominated by segments close to the interface, where mobility is enhanced but cooperative relaxation is suppressed. In contrast, low-frequency modes average over deeper regions. In the interior, high modes sample the forward moment of friction coefficients, while low modes sample the inverse moment, leading to distinct scaling behavior across the film.

## Conclusions

Since pioneering work by Plazek in the 1960's, it has been observed that polymers can exhibit anomalous chain dynamics in the glass formation range. This intersection of polymer physics (chain connectivity dominated) and glass physics (segmental motion dominated) has resisted deep understanding and posed a challenge to understanding rheological response and relaxation of glass-forming polymers. Our findings here indicate that the central mechanism by which glassy physics qualitatively alter chain dynamics is via dynamic heterogeneity. This heterogeneity-mediated scenario also unites the previously disparate problems of understanding polymer thermorheological complexity in compositionally homogeneous bulk polymers and in nanostructured polymers that are dominated by interfaces.

We find that in both bulk and near interfaces, polymer chain dynamics in the Rouse regime provide a remarkably sensitive probe of dynamic heterogeneity. Glassy dynamic heterogeneity, which grows on cooling, leads to temperature-dependent deviations from classical Rouse scaling – a consequence of probing different portions of the dynamically heterogeneous distribution with different chain modes. Perhaps counterintuitively, this has opposing effects on scaling exponents in the Rouse regime in the bulk and near free surfaces. In the bulk, the average single segment probes the forward moment of the friction coefficient distribution, which is relatively slow; higher modes interpolate towards the (faster) inverse moment. This leads to compression of the Rouse regime and higher Rouse scaling exponents.

The spatially organized nature of surface dynamic heterogeneities leads to the opposite result. Segments immediately at the surface sample the fastest-relaxing part of the friction coefficient distribution as averaged over the entire near-interface regime. Larger (slower-relaxing) modes begin to sample slower segmental dynamics deeper in the film, leading to a tether-like effect that stretches the Rouse regime and reduces its scaling exponents. Resulting Rouse scaling exponents are 'pseudo-entangled' in value, meaning that they have values similar to finite-chain entangled systems; however, they are distinguishable from topological entanglement by the lower chain displacements at which this localization is observed. This must, in turn, translate to higher surface rubbery moduli than would be expected from topological entanglement alone. It is conceivable that this observation may connect to prior reports of enhanced rubbery moduli in polymer thin films, which have thus far remained incompletely understood.[87]

Perhaps surprisingly, the surface gradient in chain dynamical scaling exponents is of qualitatively distinct form from that of the underlying segmental relaxation time gradient. The former is linear; the latter is double exponential.[14] Our results suggest, however, that the overall range of the chain mobility gradient is comparable to that of the segmental relaxation time gradient. This is consistent with an understanding wherein the segmental relaxation time gradient drives alterations in the qualitative nature and gradient form of surface chain mobility.

Taken together, these findings establish that polymer chain dynamical scaling is altered in inverse ways by bulk and surface heterogeneities, with bulk heterogeneity enhancing Rouse scaling exponents and surface heterogeneity suppressing them. This indicates that, while both bulk and interfacial polymers will exhibit breakdown of Time-Temperature-Superposition near $T_g$, the nature of the breakdown will be inverted between these two regions: bulk polymers exhibit progressive compression of the Rouse regime on cooling towards $T_g$, while near-surface polymer exhibits progressive broadening. The linear gradient in the Rouse scaling exponent $\gamma$ across depth provides a quantitative measure of the range of crossover between these regimes.

Finally, we propose that $\gamma$ serves as a sensitive and spatially resolved probe of glassy dynamic heterogeneity, capable of distinguishing between competing regimes of heterogeneity and revealing the influence of confinement and interfacial effects. This approach opens new avenues for better characterizing and understanding polymer dynamics in thin films, nanocomposites, and other confined geometries, and may inform the design of materials with tailored surface and bulk properties for applications in coatings, membranes, and flexible electronics.

**Acknowledgements**. This material is based upon work supported by the National Science Foundation under grant no. CBET – 2208238.

**Notes**. The authors declare no competing financial interest.

# References

(1) Ediger, M. D. Spatially Heterogeneous Dynamics in Supercooled Liquids. *Annual Review of Physical Chemistry* **2000**, *51* (1), 99–128. https://doi.org/10.1146/annurev.physchem.51.1.99.

(2) Cicerone, M. T.; Blackburn, F. R.; Ediger, M. D. How Do Molecules Move near Tg? Molecular Rotation of Six Probes in o-Terphenyl across 14 Decades in Time. *The Journal of Chemical Physics* **1995**, *102*, 471. https://doi.org/10.1063/1.469425.

(3) Kegel, W. K.; Blaaderen, and A. van. Direct Observation of Dynamical Heterogeneities in Colloidal Hard-Sphere

Suspensions. *Science* **2000**, *287* (5451), 290–293. https://doi.org/10.1126/science.287.5451.290.

(4) Vidal Russell, E.; Israeloff, N. E. Direct Observation of Molecular Cooperativity near the Glass Transition. *Nature* **2000**, *408* (6813), 695–698. https://doi.org/10.1038/35047037.

(5) Richert, R. Heterogeneous Dynamics in Liquids: Fluctuations in Space and Time. *J. Phys.: Condens. Matter* **2002**, *14* (23), R703–R738. https://doi.org/10.1088/0953-8984/14/23/201.

(6) Lačević, N.; Starr, F. W.; Schrøder, T. B.; Glotzer, S. C. Spatially Heterogeneous Dynamics Investigated via a Time-Dependent Four-Point Density Correlation Function. *J. Chem. Phys.* **2003**, *119* (14), 7372–7387. https://doi.org/10.1063/1.1605094.

(7) Widmer-Cooper, A.; Harrowell, P.; Fynewever, H. How Reproducible Are Dynamic Heterogeneities in a Supercooled Liquid? *Phys. Rev. Lett.* **2004**, *93* (13), 135701. https://doi.org/10.1103/PhysRevLett.93.135701.

(8) Matharoo, G. S.; Razul, M. S. G.; Poole, P. H. Structural and Dynamical Heterogeneity in a Glass-Forming Liquid. *Phys. Rev. E* **2006**, *74* (5), 050502. https://doi.org/10.1103/PhysRevE.74.050502.

(9) Widmer-Cooper, A.; Harrowell, P. Predicting the Long-Time Dynamic Heterogeneity in a Supercooled Liquid on the Basis of Short-Time Heterogeneities. *Phys. Rev. Lett.* **2006**, *96* (18), 185701. https://doi.org/10.1103/PhysRevLett.96.185701.

(10) Paeng, K.; Kaufman, L. J. Single Molecule Rotational Probing of Supercooled Liquids. *Chem. Soc. Rev.* **2014**, *43* (4), 977–989. https://doi.org/10.1039/C3CS60186B.

(11) Paeng, K.; Kaufman, L. J. Single Molecule Experiments Reveal the Dynamic Heterogeneity and Exchange Time Scales of Polystyrene near the Glass Transition. *Macromolecules* **2016**, *49* (7), 2876–2885. https://doi.org/10.1021/acs.macromol.6b00097.

(12) Paeng, K.; Kaufman, L. J. Which Probes Can Report Intrinsic Dynamic Heterogeneity of a Glass Forming Liquid? *J. Chem. Phys.* **2018**, *149* (16), 164501. https://doi.org/10.1063/1.5047215.

(13) Berthier, L. Self-Induced Heterogeneity in Deeply Supercooled Liquids. *Physical Review Letters* **2020**, *127*, 088002.

(14) Schweizer, K. S.; Simmons, D. S. Progress towards a Phenomenological Picture and Theoretical Understanding of Glassy Dynamics and Vitrification near Interfaces and under Nanoconfinement. *Journal of Chemical Physics* **2019**, *151*, 240901. https://doi.org/10.1063/1.5129405.

(15) Simmons, D. S. An Emerging Unified View of Dynamic Interphases in Polymers. *Macromol. Chem. Phys.* **2016**, *217* (2), 137–148. https://doi.org/10.1002/macp.201500284.

(16) B. Roth, C. Polymers under Nanoconfinement: Where Are We Now in Understanding Local Property Changes? *Chemical Society Reviews* **2021**, *50* (14), 8050–8066. https://doi.org/10.1039/D1CS00054C.

(17) Ediger, M. D.; Forrest, J. A. Dynamics near Free Surfaces and the Glass Transition in Thin Polymer Films: A View to the Future. *Macromolecules* **2014**, *47* (2), 471–478. https://doi.org/10.1021/ma4017696.

(18) Ellison, C. J.; Torkelson, J. M. The Distribution of Glass-Transition Temperatures in Nanoscopically Confined Glass Formers. *Nature Materials* **2003**, *2* (10), 695–700. https://doi.org/10.1038/nmat980.

(19) Baglay, R. R.; Roth, C. B. Communication: Experimentally Determined Profile of Local Glass Transition Temperature across a Glassy-Rubbery Polymer Interface with a Tg Difference of 80 K. *The Journal of Chemical Physics* **2015**, *143* (11), 111101. https://doi.org/10.1063/1.4931403.

(20) Keddie, J. L.; Jones, R. A. L.; Cory, R. A. Size-Dependent Depression of the Glass Transition Temperature in Polymer Films. *Europhysics Letters (EPL)* **1994**, *27*, 59–64. https://doi.org/10.1209/0295-5075/27/1/011.

(21) Chowdhury, M.; Guo, Y.; Wang, Y.; Merling, W. L.; Mangalara, J. H.; Simmons, D. S.; Priestley, R. D. Spatially Distributed Rheological Properties in Confined Polymers by Noncontact Shear. *J. Phys. Chem. Lett.* **2017**, *8* (6), 1229–1234. https://doi.org/10.1021/acs.jpclett.7b00214.

(22) Chai, Y.; Salez, T.; McGraw, J. D.; Benzaquen, M.; Dalnoki-Veress, K.; Raphael, E.; Forrest, J. A. A Direct Quantitative Measure of Surface Mobility in a Glassy Polymer. *Science* **2014**, *343* (6174), 994–999. https://doi.org/10.1126/science.1244845.

(23) Ghanekarade, A.; Phan, A. D.; Schweizer, K. S.; Simmons, D. S. Signature of Collective Elastic Glass Physics in Surface-Induced Long-Range Tails in Dynamical Gradients. *Nat. Phys.* **2023**, 1–7. https://doi.org/10.1038/s41567-023-01995-8.

(24) Rubinstein, M.; Colby, R. H. *Polymer Physics*; Oxford University PressOxford, 2003. https://doi.org/10.1093/oso/9780198520597.001.0001.

(25) Rouse, P. E. A Theory of the Linear Viscoelastic Properties of Dilute Solutions of Coiling Polymers. *J. Chem. Phys.* **1953**, *21* (7), 1272–1280. https://doi.org/10.1063/1.1699180.

(26) Yue, P.; Simmons, D. S. Quantitatively Connecting Experimental Time–Temperature–Superposition–Breakdown of Polymers near the Glass Transition to Dynamic Heterogeneity Via the Heterogeneous Rouse Model. *Macromolecules* **2025**, *58* (1), 109–122. https://doi.org/10.1021/acs.macromol.4c00751.

(27) Hung, J.-H.; Mangalara, J. H.; Simmons, D. S. Heterogeneous Rouse Model Predicts Polymer Chain Translational Normal Mode Decoupling. *Macromolecules* **2018**, *51* (8), 2887–2898. https://doi.org/10.1021/acs.macromol.8b00135.

(28) Plazek, D. J. Temperature Dependence of the Viscoelastic Behavior of Polystyrene. *J. Phys. Chem.* **1965**, *69* (10), 3480–3487. https://doi.org/10.1021/j100894a039.

(29) Plazek, D. J. The Temperature Dependence of the Viscoelastic Softening and Terminal Dispersions of Linear Amorphous Polymers. *J. Polym. Sci. Polym. Phys. Ed.* **1982**, *20* (4), 729–742. https://doi.org/10.1002/pol.1982.180200414.

(30) Plazek, D. L.; Plazek, D. J. Viscoelastic Behavior of Atactic Polypropylene. *Macromolecules* **1983**, *16* (9), 1469–1475. https://doi.org/10.1021/ma00243a011.

(31) McKenna, G. B.; Ngai, K. L.; Plazek, D. J. Differences in the Molecular Weight and the Temperature Dependences of Self-Diffusion and Zero Shear Viscosity in Linear Polyethylene and Hydrogenated Polybutadiene. *Polymer* **1985**, *26* (11), 1651–1653. https://doi.org/10.1016/0032-3861(85)90280-0.

(32) Ngai, K. L.; Plazek, D. J. A Quantitative Explanation of the Difference in the Temperature Dependences of the Viscoelastic Softening and Terminal Dispersions of Linear Amorphous Polymers. *J. Polym. Sci. B Polym. Phys.* **1986**, *24* (3), 619–632. https://doi.org/10.1002/polb.1986.090240310.

(33) Plazek, D. J.; Zheng, X. D.; Ngai, K. L. Viscoelastic Properties of Amorphous Polymers. I. Different Temperature Dependences of Segmental Relaxation and Terminal Dispersion. *Macromolecules* **1992**, *25* (19), 4920–4924. https://doi.org/10.1021/ma00045a016.

(34) Plazek, D. J.; Bero, C. A.; Neumeister, S.; Floudas, G.; Fytas, G.; Ngai, K. L. Viscoelastic Properties of Amorphous Polymers 3: Low Molecular Weight Poly(Methylphenylsiloxane). *Colloid Polym Sci* **1994**, *272* (11), 1430–1438. https://doi.org/10.1007/BF00654173.

(35) Plazek, D. J.; Chay, I.-C.; Ngai, K. L.; Roland, C. M. Viscoelastic Properties of Polymers. 4. Thermorheological Complexity of the Softening Dispersion in Polyisobutylene. *Macromolecules* **1995**, *28* (19), 6432–6436. https://doi.org/10.1021/ma00123a007.

(36) Sokolov, A. P.; Schweizer, K. S. Resolving the Mystery of the Chain Friction Mechanism in Polymer Liquids. *Phys. Rev. Lett.* **2009**, *102* (24), 248301. https://doi.org/10.1103/PhysRevLett.102.248301.

(37) Agapov, A. L.; Novikov, V. N.; Hong, T.; Fan, F.; Sokolov, A. P. Surprising Temperature Scaling of Viscoelastic Properties in Polymers. *Macromolecules* **2018**, *51* (13), 4874–4881. https://doi.org/10.1021/acs.macromol.8b00454.

(38) Sokolov, A. P.; Hayashi, Y. Breakdown of Time–Temperature Superposition: From Experiment to the Coupling Model and Beyond. *Journal of Non-Crystalline Solids* **2007**, *353* (41–43), 3838–3844. https://doi.org/10.1016/j.jnoncrysol.2007.02.063.

(39) Ding, Y.; Sokolov, A. P. Breakdown of Time−Temperature Superposition Principle and Universality of Chain Dynamics in Polymers. *Macromolecules* **2006**, *39* (9), 3322–3326. https://doi.org/10.1021/ma052607b.

(40) He, Y.; Lutz, T. R.; Ediger, M. D.; Ayyagari, C.; Bedrov, D.; Smith, G. D. NMR Experiments and Molecular Dynamics Simulations of the Segmental Dynamics of Polystyrene. *Macromolecules* **2004**, *37* (13), 5032–5039. https://doi.org/10.1021/ma049843r.

(41) Zorn, R.; McKenna, G. B.; Willner, L.; Richter, D. Rheological Investigation of Polybutadienes Having Different Microstructures over a Large Temperature Range. *Macromolecules* **1995**, *28* (25), 8552–8562. https://doi.org/10.1021/ma00129a014.

(42) Ngai, K. L.; Plazek, D. J.; Roland, C. M. Comment on ``Resolving the Mystery of the Chain Friction Mechanism in Polymer Liquids''. *Phys. Rev. Lett.* **2009**, *103* (15), 159801. https://doi.org/10.1103/PhysRevLett.103.159801.

(43) Sokolov, A. P.; Schweizer, K. S. Resolving the Mystery of the Chain Friction Mechanism in Polymer Liquids. *Phys. Rev. Lett.* **2009**, *102* (24), 248301. https://doi.org/10.1103/PhysRevLett.102.248301.

(44) Sokolov, A. P.; Schweizer, K. S. Sokolov and Schweizer Reply: *Phys. Rev. Lett.* **2009**, *103* (15), 159802. https://doi.org/10.1103/PhysRevLett.103.159802.

(45) Inoue, T.; Onogi, T.; Yao, M.-L.; Osaki, K. Viscoelasticity of Low Molecular Weight Polystyrene. Separation of Rubbery and Glassy Components. *J. Polym. Sci. B Polym. Phys.* **1999**, *37* (4), 389–397. https://doi.org/10.1002/(SICI)1099-0488(19990215)37:4%253C389::AID-POLB12%253E3.0.CO;2-G.

(46) Ghanekarade, A.; Phan, A. D.; Schweizer, K. S.; Simmons, D. S. Nature of Dynamic Gradients, Glass Formation, and Collective Effects in Ultrathin Freestanding Films. *PNAS* **2021**, *118* (31), e2104398118. https://doi.org/10.1073/pnas.2104398118.

(47) Phan, A. D.; Schweizer, K. S. Dynamic Gradients, Mobile Layers, Tg Shifts, Role of Vitrification Criterion, and Inhomogeneous Decoupling in Free-Standing Polymer

Films. *Macromolecules* **2018**, *51* (15), 6063–6075. https://doi.org/10.1021/acs.macromol.8b01094.

(48) Phan, A. D.; Schweizer, K. S. Theory of Spatial Gradients of Relaxation, Vitrification Temperature and Fragility of Glass-Forming Polymer Liquids Near Solid Substrates. *ACS Macro Lett.* **2020**, *9* (4), 448–453. https://doi.org/10.1021/acsmacrolett.0c00006.

(49) Phan, A. D.; Schweizer, K. S. Influence of Longer Range Transfer of Vapor Interface Modified Caging Constraints on the Spatially Heterogeneous Dynamics of Glass-Forming Liquids. *Macromolecules* **2019**, *52* (14), 5192–5206. https://doi.org/10.1021/acs.macromol.9b00754.

(50) Ghanekarade, A.; Phan, A. D.; Schweizer, K. S.; Simmons, D. S. Signature of Collective Elastic Glass Physics in Surface-Induced Long-Range Tails in Dynamical Gradients. *Nat. Phys.* **2023**, *19*, 800–806. https://doi.org/10.1038/s41567-023-01995-8.

(51) Lam, C.-H.; Tsui, O. K. C. Crossover to Surface Flow in Supercooled Unentangled Polymer Films. *Phys. Rev. E* **2013**, *88* (4), 042604. https://doi.org/10.1103/PhysRevE.88.042604.

(52) Hao, Z.; Ghanekarade, A.; Zhu, N.; Randazzo, K.; Kawaguchi, D.; Tanaka, K.; Wang, X.; Simmons, D. S.; Priestley, R. D.; Zuo, B. Mobility Gradients Yield Rubbery Surfaces on Top of Polymer Glasses. *Nature* **2021**, *596* (7872), 372–376. https://doi.org/10.1038/s41586-021-03733-7.

(53) Plimpton, S. Fast Parallel Algorithms for Short-Range Molecular Dynamics. *Journal of Computational Physics* **1995**, *117* (1), 1–19. https://doi.org/10.1006/jcph.1995.1039.

(54) Grest, G. S.; Kremer, K. Molecular Dynamics Simulation for Polymers in the Presence of a Heat Bath. *Physical Review A* **1986**, *33* (5), 3628.

(55) Baschnagel, J.; Varnik, F. Computer Simulations of Supercooled Polymer Melts in the Bulk and in Confined Geometry. *J. Phys.: Condens. Matter* **2005**, *17* (32), R851. https://doi.org/10.1088/0953-8984/17/32/R02.

(56) Kremer, K.; Grest, G. S. Dynamics of Entangled Linear Polymer Melts: A Molecular-dynamics Simulation. *J. Chem. Phys.* **1990**, *92* (8), 5057–5086. https://doi.org/10.1063/1.458541.

(57) Mackura, M. E.; Simmons, D. S. Enhancing Heterogenous Crystallization Resistance in a Bead-Spring Polymer Model by Modifying Bond Length. *J. Polym. Sci. Part B: Polym. Phys.* **2014**, *52* (2), 134–140. https://doi.org/10.1002/polb.23398.

(58) Vela, D.; Ghanekarade, A.; Simmons, D. S. Probing the Metrology and Chemistry Dependences of the Onset Condition of Strong "Nanoconfinement" Effects on Dynamics. *Macromolecules* **2020**, *Article ASAP*. https://doi.org/10.1021/acs.macromol.9b02693.

(59) Diaz-Vela, D.; Hung, J.-H.; Simmons, D. S. Temperature-Independent Rescaling of the Local Activation Barrier Drives Free Surface Nanoconfinement Effects on Segmental-Scale Translational Dynamics near Tg. *ACS Macro Lett.* **2018**, *7* (11), 1295–1301. https://doi.org/10.1021/acsmacrolett.8b00695.

(60) Hung, J.-H.; Patra, T. K.; Meenakshisundaram, V.; Mangalara, J. H.; Simmons, D. S. Universal Localization Transition Accompanying Glass Formation: Insights from Efficient Molecular Dynamics Simulations of Diverse Supercooled Liquids. *Soft Matter* **2019**, *15* (6), 1223–1242. https://doi.org/10.1039/C8SM02051E.

(61) Peter, S.; Meyer, H.; Baschnagel, J. Thickness-dependent Reduction of the Glass-transition Temperature in Thin Polymer Films with a Free Surface. *Journal of Polymer Science Part B: Polymer Physics* **2006**, *44* (20), 2951–2967. https://doi.org/10.1002/polb.20924.

(62) Hanakata, P. Z.; Douglas, J. F.; Starr, F. W. Interfacial Mobility Scale Determines the Scale of Collective Motion and Relaxation Rate in Polymer Films. *Nat Commun* **2014**, *5*, 4163. https://doi.org/10.1038/ncomms5163.

(63) Hanakata, P. Z.; Betancourt, B. A. P.; Douglas, J. F.; Starr, F. W. A Unifying Framework to Quantify the Effects of Substrate Interactions, Stiffness, and Roughness on the Dynamics of Thin Supported Polymer Films. *The Journal of Chemical Physics* **2015**, *142* (23), 234907. https://doi.org/10.1063/1.4922481.

(64) Lang, R. J.; Merling, W. L.; Simmons, D. S. Combined Dependence of Nanoconfined Tg on Interfacial Energy and Softness of Confinement. *ACS Macro Lett.* **2014**, *3* (8), 758–762. https://doi.org/10.1021/mz500361v.

(65) Riggleman, R. A.; Yoshimoto, K.; Douglas, J. F.; de Pablo, J. J. Influence of Confinement on the Fragility of Antiplasticized and Pure Polymer Films. *Physical Review Letters* **2006**, *97* (4), 045502.

(66) Riggleman, R. A.; Douglas, J. F.; de Pablo, J. J. Antiplasticization and the Elastic Properties of Glass-Forming Polymer Liquids. *Soft Matter* **2010**, *6*, 292–304.

(67) Starr, F. W.; Douglas, J. F.; Meng, D.; Kumar, S. K. Bound Layers "Cloak" Nanoparticles in Strongly Interacting Polymer Nanocomposites. *ACS Nano* **2016**, *10* (12), 10960–10965. https://doi.org/10.1021/acsnano.6b05683.

(68) Varnik, F.; Baschnagel, J.; Binder, K. Reduction of the Glass Transition Temperature in Polymer Films: A Molecular-Dynamics Study. *Phys. Rev. E* **2002**, *65* (2), 021507. https://doi.org/10.1103/PhysRevE.65.021507.

(69) Rahman, T.; Simmons, D. S. Near-Substrate Gradients in Chain Relaxation and Viscosity in a Model Low-Molecular Weight Polymer. *Macromolecules* **2021**, *54* (13), 5935–5949. https://doi.org/10.1021/acs.macromol.0c02888.

(70) AMDAT: Amorphous Molecular Dynamics Analysis Toolkit. 10.5281/zenodo.17417166.

(71) Kawak, P.; Drayer, W. F.; Simmons, D. S. AMDAT: An Open-Source Molecular Dynamics Analysis Toolkit for Supercooled Liquids, Glass-Forming Materials, and Complex Fluids. arXiv February 5, 2026. https://doi.org/10.48550/arXiv.2602.05865.

(72) Binder, K.; Baschnagel, J.; Paul, W. Glass Transition of Polymer Melts: Test of Theoretical Concepts by Computer Simulation. *Progress in Polymer Science* **2003**, *28* (1), 115–172. https://doi.org/10.1016/S0079-6700(02)00030-8.

(73) Schweizer, K. S.; Saltzman, E. J. Entropic Barriers, Activated Hopping, and the Glass Transition in Colloidal Suspensions. *The Journal of Chemical Physics* **2003**, *119* (2), 1181–1196. https://doi.org/10.1063/1.1578632.

(74) Kremer, K.; Grest, G. S. Dynamics of Entangled Linear Polymer Melts: A Molecular-Dynamics Simulation. *The Journal of Chemical Physics* **1990**, *92* (8), 5057–5086. https://doi.org/10.1063/1.458541.

(75) Kob, W.; Andersen, H. C. Testing Mode-Coupling Theory for a Supercooled Binary Lennard-Jones Mixture I: The van Hove Correlation Function. *Phys. Rev. E* **1995**, *51* (5), 4626–4641. https://doi.org/10.1103/PhysRevE.51.4626.

(76) Grest, G. S. Communication: Polymer Entanglement Dynamics: Role of Attractive Interactions. *Journal of Chemical Physics* **2016**, *145* (14). https://doi.org/10.1063/1.4964617.

(77) Hanakata, P. Z.; Douglas, J. F.; Starr, F. W. Local Variation of Fragility and Glass Transition Temperature of Ultra-Thin Supported Polymer Films. *The Journal of Chemical Physics* **2012**, *137* (24), 244901. https://doi.org/doi:10.1063/1.4772402.

(78) Lang, R. J.; Simmons, D. S. Interfacial Dynamic Length Scales in the Glass Transition of a Model Freestanding Polymer Film and Their Connection to Cooperative Motion. *Macromolecules* **2013**, *46* (24), 9818–9825. https://doi.org/10.1021/ma401525q.

(79) DeFelice, J.; Milner, S. T.; Lipson, J. E. G. Simulating Local Tg Reporting Layers in Glassy Thin Films. *Macromolecules* **2016**, *49* (5), 1822–1833. https://doi.org/10.1021/acs.macromol.6b00090.

(80) Sussman, D. M.; Tung, W.-S.; Winey, K. I.; Schweizer, K. S.; Riggleman, R. A. Entanglement Reduction and Anisotropic Chain and Primitive Path Conformations in Polymer Melts under Thin Film and Cylindrical Confinement. *Macromolecules* **2014**, *47* (18), 6462–6472. https://doi.org/10.1021/ma501193f.

(81) Hao, Z.; Ghanekarade, A.; Zhu, N.; Randazzo, K.; Kawaguchi, D.; Tanaka, K.; Wang, X.; Simmons, D. S.; Priestley, R. D.; Zuo, B. Mobility Gradients Yield Rubbery Surfaces on Top of Polymer Glasses. *Nature* **2021**, *596* (7872), 372–376. https://doi.org/10.1038/s41586-021-03733-7.

(82) SciPy Developers. Scipy.Interpolate.UnivariateSpline — SciPy v1.16.1 Manual. https://docs.scipy.org/doc/scipy/reference/generated/scipy.interpolate.UnivariateSpline.html (accessed 2026-01-29).

(83) Dierckx, P. An Algorithm for Smoothing, Differentiation and Integration of Experimental Data Using Spline Functions. *Journal of Computational and Applied Mathematics* **1975**, *1* (3), 165–184. https://doi.org/10.1016/0771-050X(75)90034-0.

(84) Dierckx, P. A Fast Algorithm for Smoothing Data on a Rectangular Grid While Using Spline Functions. *SIAM J. Numer. Anal.* **1982**, *19* (6), 1286–1304. https://doi.org/10.1137/0719093.

(85) Dierckx, P. *An Improved Algorithm for Curve Fitting with Spline Functions*; TW54; Department of Computer Science, K.U. Leuven: Leuven, Belgium, 1981.

(86) Dierckx, P. *Curve and Surface Fitting with Splines*; Oxford University PressOxford, 1993. https://doi.org/10.1093/oso/9780198534419.001.0001.

(87) Li, X.; McKenna, G. B. Ultrathin Polymer Films: Rubbery Stiffening, Fragility, and Tg Reduction. *Macromolecules* **2015**, *48* (17), 6329–6336. https://doi.org/10.1021/acs.macromol.5b01263.